\documentclass{iaus}
\usepackage{graphicx}
\usepackage{natbib}
\usepackage{makeidx}

\def\note #1]{{\bf #1]}}
\def\K{\,{\rm K}}
\def\g{\,{\rm g}}
\def\s{\,{\rm s}}
\def\erg{\,{\rm erg}}
\def\cm{\,{\rm cm}}
\def\yr{\,{\rm yr}}
\def\Gyr{\,{\rm Gyr}}
\def\rt{r_{\rm t}}

\usepackage{index}
\newindex{author}{idx}{ind}{Author Index}
\newindex{object}{odx}{ond}{Object Index}
\newindex{subject}{sdx}{snd}{Subject Index}
\makeindex

\title[Solar calibration] %% give here short title %%
{The Sun as a fundamental calibrator of stellar evolution}

\author[J{\o}rgen Christensen-Dalsgaard]   %% give here short author list %%
{J{\o}rgen Christensen-Dalsgaard}
\index[author]{Christensen-Dalsgaard, J.}

\affiliation{Danish Asteroseismology Centre, 
and Department of Physics and Astronomy, \\
Aarhus University, DK 8000 Aarhus C, Denmark
\\ email: {\tt jcd@phys.au.dk} 
}

\pubyear{2009}
\volume{258}  
\pagerange{429--440}
\jname{The Ages of Stars}
\editors{E.E. Mamajek, D.R. Soderblom \& R.F.G. Wyse, eds.}
\begin{document}

\maketitle

\begin{abstract} The Sun is unique amongst stars in having a precisely
determined age which does not depend on the modelling of stellar
evolution.  Furthermore, other global properties of the Sun are known
to much higher accuracy than for any other star.  Also,
helioseismology has provided detailed determination of the solar
internal structure and rotation.  As a result, the Sun plays a central
role in the development and test of stellar modelling.  Here I discuss
solar modelling and its application to tests of asteroseismic
techniques for stellar age determination.
\index[subject]{Sun: evolution}
\index[subject]{Sun: interior}
\index[subject]{Sun: helioseismology}
\index[subject]{stars: evolution}
\index[subject]{stars: interior}
\index[subject]{stars: oscillations}
\index[subject]{stars: fundamental parameters}
\end{abstract}

\firstsection

\section{Introduction}

Determining the age of a star from its observed properties requires a
model describing how those properties change as the star evolves.
Amongst the many properties discussed in this volume, those relevant
to using the Sun as a calibrator depend on the changes in the internal
structure of the star caused by evolution.  Specifically, the change
in the composition as hydrogen is fused to helium (with the additional
effects of diffusion and settling) changes the internal structure and
hence the observable properties.  Evidently, both the details of the
composition change, and the response of the structure and the
observables, depend on the modelling of stellar interiors, and hence
the age determination is sensitive to uncertainties in the modelling.
The Sun provides a unique possibility for quantifying these
uncertainties and attempting to reduce them.

%\note [Note on the very special case of the Sun: close enough to study in 
%great detail; and the only real case where the age is known independently.]

In the present case the Sun has several major advantages: it is the
only star for which the age can be inferred in a manner that is
essentially independent of stellar modelling, through the radioactive
dating of the solar system.  Also, its proximity means that its mass,
radius and luminosity are known to high accuracy.  Finally, as a
result of this proximity, helioseismology based on a broad range of
modes has allowed inferences to be made of the detailed internal
structure (and rotation) of the Sun.  This allows a refined test of
the modelling of stellar evolution.  In addition, by applying
techniques for age determination, based on stellar modelling, to the
Sun we can test and possibly improve them.  This, in particular,
applies to the use of asteroseismic analyses for age determination; as
discussed by Lebreton \& Montalb\'an (this volume) these promise to be
far more precise than other techniques based on the evolution of
stellar structure.

%\note [Hence use the Sun to test age determination procedures, before
%applying them to other stars.]
%
%\note [Somewhere, but possibly not here, discuss problems with stellar 
%modelling.]

\section{Solar modelling}

%\note [Brief reminder of what is known about the Sun, and how.
%Possibly already here mention abundance problems.]
%
A summary of solar modelling and the helioseismic investigations of
the Sun was given, for example, by \citet{Christ2002}.
\citet{Bahcal2006} made a detailed analysis of the sensitivity
of the models to the choice of input parameters and physics.
The solar models used here largely correspond to the so-called Model S of
\citet{Christ1996}, although with some updates.

The mass of the Sun is known from planetary motion, with an accuracy
limited by the accuracy of the determination of the gravitational
constant $G$.  For the modelling presented here I use $G = 6.67232
\times 10^{-8}$ in {\it cgs} units, and hence the solar mass is
$M_\odot = 1.989 \times 10^{33} \g$.  The solar radius $R_\odot$ is
obtained from the solar angular diameter and the distance to the Sun;
the radius should be defined in a manner that can be related precisely
to the model, e.g., as the distance from the centre to the
photosphere, defined by the location where the temperature is equal to
the effective temperature.  A commonly used value is $R_\odot = 6.9599
\times 10^{10} \cm$ \citep{Auwers1891}, and this value is used here.
However, I note that \citet{Brown1998} obtained the value $6.9551
\times 10^{10} \cm$ which is probably more accurate.  The solar
luminosity $L_\odot$ is obtained from the solar constant, i.e., solar
flux at the Earth, and the distance to the Sun; this assumes that the
solar flux does not depend on latitude, a probably reasonable
assumption although one that has never been verified; I use the value
$L_\odot = 3.846 \times 10^{33} \erg \s^{-1}$
\citep[see also][]{Frohli2004}.

The solar photospheric composition can be determined from spectral
analysis.  An important exception is the abundance of helium: the
helium lines that led to the detection of helium in the Sun (and hence
the name of the element) are formed in the chromosphere, under
conditions such that the abundance determination is quite inaccurate.
The abundances of other noble gases, amongst which neon is a
relatively important constituent of the solar atmosphere, are
similarly inaccurate.
%\note [All this might deserve a little check, and a reference or two; 
%possibly check ISSI book.]
Thus only relative abundances, commonly defined relative to hydrogen,
can be determined.  In solar modelling this is typically characterized
by the ratio $Z_{\rm s}/X_{\rm s}$ of the surface abundances by mass
$Z_{\rm s}$ of elements heavier than helium and $X_{\rm s}$ of
hydrogen.  A commonly used value has been $Z_{\rm s}/X_{\rm s} =
0.0245$ \citep{Greves1993}.  However, redeterminations of the
abundances, based on three-dimensional hydrodynamical models of the
solar atmosphere and taking departures from local thermodynamical
equilibrium into account, have had a major impact on the inferred
abundances; in particular, the abundances of oxygen, carbon and
nitrogen were substantially reduced
\citep[e.g.,][]{Asplun2004} \citep[for a review, see also][]{Asplun2005}.
This resulted in $Z_{\rm s}/X_{\rm s} = 0.0165$;
as discussed below, this has had drastic consequences for the comparison
between solar models and the helioseismically inferred structure.

%\note [Discuss age in a little detail. Need to check Wasserburg again.
%Wasserburg, in appendix of \citet{Bahcal1995}:
%age at `very earliest phase of the
%main-sequence behavior' is between $4.563 \times 10^9 \yr$
%and $4.576 \times 10^9 \yr$.]

For the present discussion the solar age is of course of central
importance.  This can be inferred from radioactive dating of material
from the early solar system, as represented by suitable meteorites.
In a detailed discussion presented in the appendix to
\citet{Bahcal1995}, G. J. Wasserburg concluded that the age of the Sun, since the beginning of its
main-sequence evolution, is between $4.563 \times 10^9$ and $4.576
\times 10^9 \yr$.  The rounded value of $4.6 \times 10^9 \yr$ is often
used, including for the reference Model S.

%\note [Calibration of solar models. Could also mention here tests of
%determination of alpha etc. from 3D simulations.]

It is evident that computed solar models should match the observed quantities,
at the age of the Sun.
Models are typically computed without mass loss
\citep[see, however,][]{Sackma2003} and hence with the present mass of the Sun.
A model with the correct radius, luminosity and $Z_{\rm s}/X_{\rm s}$
is obtained by adjusting three parameters that are {\it a priori\/}
unknown: a parameter, such as the mixing-length parameter $\alpha_{\rm
ML}$, \footnote{i.e., the ratio of the mixing length to the pressure
scale height in the mixing-length description of convection
\citep[e.g.,][]{Bohm1958, Gough1976}} 
characterizing the properties of convection
which largely determines the radius, the initial helium abundance $Y_0$
which mainly determines the luminosity, and the initial heavy-element
abundance $Z_0$ which determines $Z_{\rm s}/X_{\rm s}$.
This calibration provides a precise (although not necessarily accurate)
determination of the initial solar helium abundance, of importance to studies
of galactic chemical evolution.
Also, the resulting value of $\alpha_{\rm ML}$ is often used for computations
of other stellar models, although there is little justification for
regarding $\alpha_{\rm ML}$ as being independent of stellar parameters.
More significantly, the calibration of the properties of the solar convection
zone can be used as a test of hydrodynamical simulations of convection
\citep[e.g.,][]{Demarq1999, Rosent1999},
these may then be used to calibrate the dependence of 
convection-zone properties, e.g., characterized by $\alpha_{\rm LM}$,
on stellar parameters \citep{Ludwig1997, Ludwig1999, Trampe1999}.

An important solar observable is the neutrino flux.  The discrepancy
between the predicted and detected flux of electron neutrinos was long
regarded as a potential problem of solar modelling, although even
early helioseismic results strongly indicated that changes to the
models designed to eliminate the discrepancy were inconsistent with
the observed oscillation frequencies \citep[e.g.,][]{Elswor1990}.
However, it is now realized that the apparent discrepancy was caused
by oscillations between different states of the neutrino; with the
recent detection of neutrinos of other flavours the total observed
flux of neutrinos agrees with predictions \citep{Ahmad2002}.  Thus the
emphasis in the study of solar neutrinos has shifted towards the
investigation of the detailed properties of the neutrino oscillations
\citep[for recent reviews, see][]{Bahcal2004, Robert2006, Haxton2006, 
Haxton2008}.  Interestingly, the computed neutrino flux is not
significantly affected by the recent revision in the solar composition
\citep[e.g.][]{Bahcal2005a}.
%\note [with a recent reference. And we need to think, also for the book(!),
%about the diagnostic potential of this.]

\section{Helioseismic results on the solar interior}

%\note [Brief summary of observations, possibly mention near-surface errors.]
%
Very extensive data on solar oscillations have been obtained in last
two decades \citep[for a review, see][]{Christ2002}.  Unlike any other
pulsating star the availability of observations with high spatial
resolution has provided accurate frequencies for modes over a broad
range of spherical-harmonic degrees $l$, from 0 to more than 1000 (see
also Lebreton \& Montalb\'an, these proceedings, for an overview of
the properties of stellar oscillations).  Most of the observed modes
are acoustic modes; these are essentially trapped between the solar
surface and an inner {\it turning point\/} at a distance $\rt$ from
the centre given by $c(\rt)/\rt = \omega/\sqrt{l(l+1)}$, where $c$ is
the adiabatic sound speed and $\omega$ is the angular frequency of the
mode.  Thus the broad range of $l$ corresponds to a range of inner
turning points varying from the centre to just beneath the solar
surface.  This availability of modes sensitive to very different parts
of the Sun is essentially what allows inverse analyses to resolve the
structure and rotation of the solar interior.

In the analysis of solar and solar-like pulsations an important issue
is the effect of the near-surface layers: modelling of the structure
of these layers and of their effect on the oscillation frequencies is
highly uncertain, leading to systematic errors in the computed
frequencies, which in many cases dominate the differences between
observed and computed solar frequencies \citep[e.g.,][]{Christ1988a}.
These errors depend essentially only on frequency (apart from a
trivial dependence on the mode inertia) and furthermore are small at
low frequency \citep[e.g.,][]{Christ1997}.  This allows their effect
to be eliminated in the analysis of solar data but they should be kept
in mind also in asteroseismic analyses \citep{Kjelds2008a}.

%\note [Briefly on results of solar inversions, probably include also
%models with revised composition.
%Note that this might give evidence of basic problems in stellar modelling.]

\begin{figure}[b]
 %\vspace*{-0.5 cm}
\begin{center}
 \includegraphics[width=9cm]{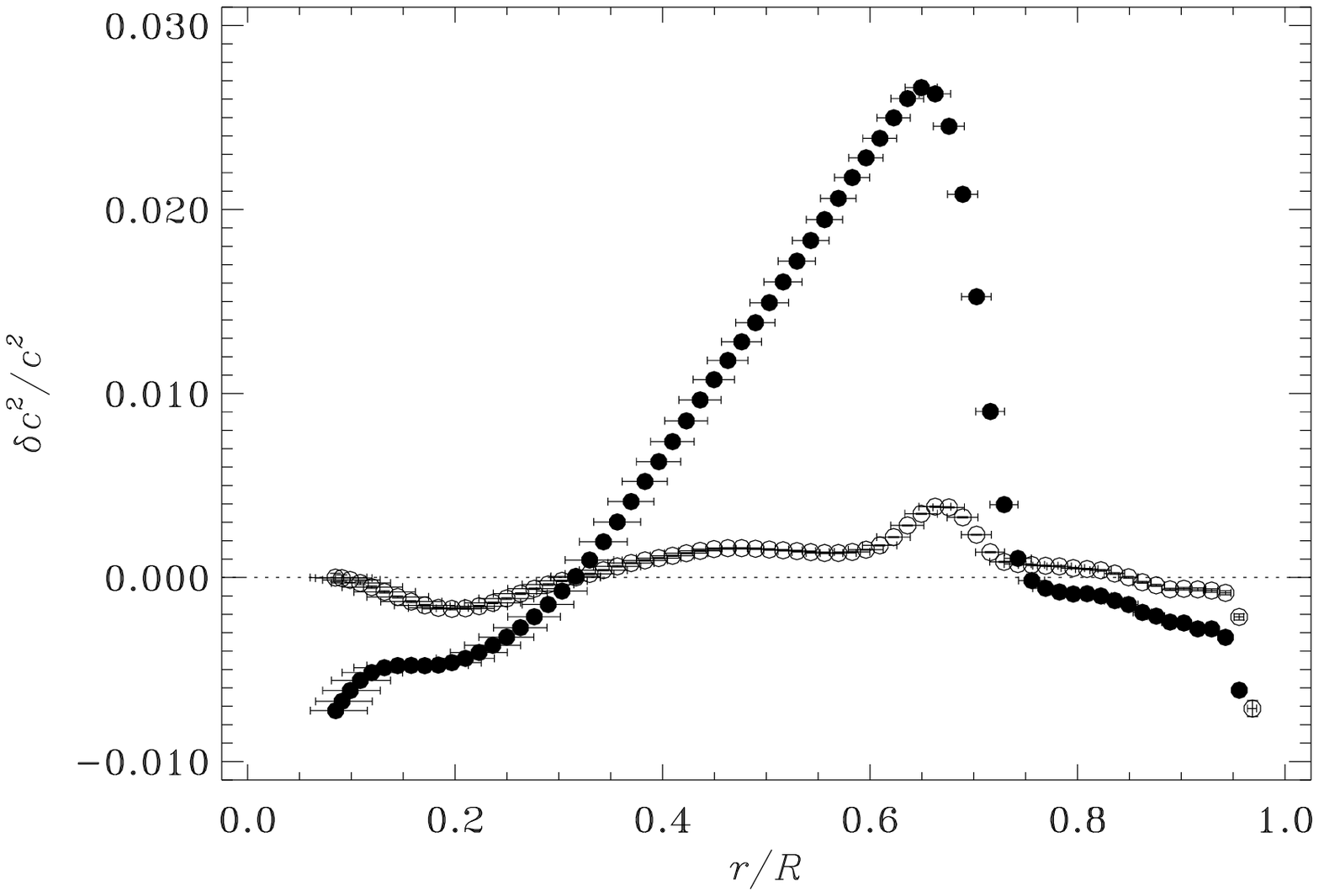} 
% \vspace*{-1.0 cm}
 \caption{Inferred relative differences in squared sound speed, in the
sense (Sun) -- (model), from inversion of frequency differences between the
Sun and two solar models.
The open circles used Model S of \citet{Christ1996},
based on the \citet{Greves1993} composition, while the filled
circles are for a corresponding model but using the revised solar
composition \citep{Asplun2005}.
The horizontal bars provide a measure of the resolution of the inversion.
The standard error in the inferred differences are generally smaller than
the size of the symbols.
\citep[From][]{Christ2009}.}
   \label{fig:cinv}
\end{center}
\end{figure}

To illustrate the inference of solar structure, Fig.~\ref{fig:cinv}
shows the inferred difference in squared sound speed between the Sun 
and two solar models.
One is what might be termed a `standard' model,
computed with the \citet{Greves1993} composition.
Here the relative differences are below 0.5\%; although this is far more than
the very small estimated errors in the difference it still indicates that
the model provides a good representation of the solar interior.
Also shown are the results for a corresponding model computed with the revised
\citep{Asplun2005} composition.
It is evident that this leads to a dramatic deterioration in the
agreement between the model and the Sun
\citep[see also][for a review]{Basu2008}.
Similar discrepancies are found for other helioseismically inferred quantities,
such as the depth of the convection zone and the envelope helium abundance.
Of particular relevance to the present discussion is the inconsistency 
found in a detailed analysis by \citet{Chapli2007} in
the small frequency separation
$\delta \nu_{l\,l+2}(n) = \nu_{nl} - \nu_{n-1\,l+2}$,
$\nu_{nl}$ being the cyclic frequency of a mode of degree $l$ and radial
order $n$;
as discussed by Lebreton \& Montalb\'an (this volume) this is
a measure of stellar age.
I return to the consequences of this below.

These discrepancies clearly indicate potential problems with solar modelling,
if the revision of the solar abundances is accepted.
\citet{Guzik2006, Guzik2008} reviewed the attempts to modify the model
calculation to restore the agreement with helioseismology which so far have
not led to any entirely satisfactory solution.
A trivial modification is to postulate intrinsic errors in the 
opacity tables which compensate for the composition change
\citep{Bahcal2005b, Christ2009};
however, as discussed in the latter reference the required change is as high 
as 30\% at the base of the convection zone, at a temperature of
$2 \times 10^6 \K$, which may be unrealistic.
A resolution of these issues is evidently of general importance to stellar
modelling and hence to the determination of stellar ages from evolution
calculations.

%\note [Also, mention inferences of rotation, and relevance for gyrochronology
%(we need to understand the evolution of the rotation rate, including transport
%in the interior, to get a theoretical verification of the method).]

I finally recall that helioseismology has yielded detailed inferences of
the solar internal rotation \citep[see][for a review]{Thomps2003}.
This shows that the convection zone approximately shares the surface
latitudinal differential rotation, while the radiative interior rotates
at a nearly constant rate, somewhat smaller than the surface equatorial
rotation rate.
This is obviously relevant to the modelling of the,
so far incompletely understood, evolution of stellar rotation and
hence to the use of gyrochronology for stellar age determinations
\citep[][Meibom, this volume]{Barnes2007}.

\begin{figure}[b]
% \vspace*{-2.0 cm}
\begin{center}
 \includegraphics[width=13cm]{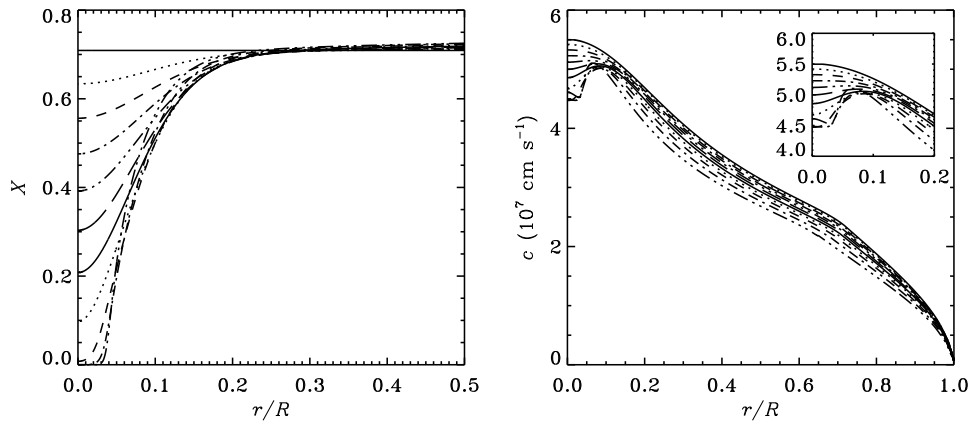} 
% \vspace*{-1.0 cm}
 \caption{The left-hand panel shows the hydrogen abundance $X$ in the 
 inner part of $1 \, M_\odot$ models of age $0 - 10 \Gyr$, in steps
 of $1 \Gyr$.
 The right-hand panel shows the sound speed in these models,
 with an enlargement in the insert of the behaviour in the core.
 Except for the final model the central sound speed decreases with
 increasing age.}
   \label{fig:cevol}
\end{center}
\end{figure}

\section{Asteroseismic age determination}

%\note [Summarize evolution, including composition change.]
%
As discussed by Lebreton \& Montalb\'an (this volume), asteroseismology%
\footnote{For an illuminating and entertaining discussion of the etymology
of {\it asteroseismology\/}, see \citet{Gough1996}.}
provides sensitive diagnostics of stellar ages.
The change in the internal structure of a star with evolution directly
affects the oscillation frequencies,
and hence the observed frequencies, when suitably analyzed, can be used to
determine the age.
Also, the computation of the relevant aspects of the frequencies from a
given model structure is relatively insensitive to systematic errors;
on the other hand, the dependence of the structure on age is clearly
affected by uncertainties in the modelling.

%\note [Summarize resulting sound-speed change.]

Here I concentrate on acoustic modes in solar-like stars;
these are typically of high radial order and hence their diagnostic potential 
can be investigated on the basis of asymptotic theory.
The low-degree modes that are relevant to observations of distant stars
penetrate to the core of the star; 
hence their frequencies are sensitive to the sound speed in the core
and consequently to the composition change resulting from evolution.
In the ideal-gas approximation $c^2 \propto T/\mu$, where $T$ is
temperature and $\mu$ is the mean molecular weight.
Both increase as a result of evolution;
however, since the temperature is strongly constrained by the high
temperature sensitivity of the nuclear burning rates, the change in 
$\mu$ dominates, leading to a decrease in the sound speed in the core.
This is illustrated in Fig.~\ref{fig:cevol} for the evolution of 
a $1 \, M_\odot$ star.
The decrease in $c$ affects most strongly the modes of the lowest
degree which penetrate most deeply;
consequently, the small frequency separations $\delta \nu_{02}$ and
$\delta \nu_{13}$ decrease with increasing age. 
To characterize a star based on high-order acoustic-mode frequencies one can
in addition use the {\it large frequency separation\/}
$\Delta \nu_{nl} = \nu_{nl} - \nu_{n-1\,l}$ which essentially provides a
measure of the mean density of the star.
Thus the position of the star in a $(\langle \Delta\nu \rangle,
\langle \delta \nu \rangle)$ diagram, based on suitable averages, 
provides an indication of the mass and age of the star
\citep{Christ1984, Christ1988b, Ulrich1986}.
Obviously, the calibration of the diagram depends on the physics and
other parameters, such as the composition, of the stars
\citep[][see also Lebreton \& Montalb\'an, these proceedings]
{Gough1987, Montei2002}.
These potential systematic errors must be taken into account in 
the interpretation of the results.

It was noted by \citet{Roxbur2003} that the near-surface errors, although
to a large extent canceling in the difference, has a significant effect
on the small frequency separations.
They showed that this effect is suppressed by evaluating separation
ratios, such as
\begin{equation}
r_{02}(n) = {\nu_{n0} - \nu_{n-1\,2} \over \nu_{n1} - \nu_{n-1\,1}} \; , \qquad
r_{13}(n) = {\nu_{n1} - \nu_{n-1\,3} \over \nu_{n+1\,0} - \nu_{n0}} \; ,
\label{eq:ratio}
\end{equation}
and demonstrated that these ratios are directly related to the effect of the
stellar core on the oscillation frequencies.
This was further analyzed by \citet{OtiFlo2005} who showed that, unlike
$\delta \nu_{02}$, $r_{02}$ is essentially insensitive to structure changes
near the surface;
they furthermore considered several examples of model modifications,
including changes to the stellar radius, keeping the structure of the core
unchanged and found that these did not affect the separation ratios.
The use of $(\langle \Delta \nu \rangle, \langle r_{02} \rangle)$
diagrams to characterize stellar properties is discussed by
Lebreton \& Montalb\'an (this volume).

This correction for the near-surface effects assumes that they are
independent of degree and hence essentially that the underlying
physical cause is spherically symmetric.
In fact, it is known from the case of the Sun that the magnetic
activity causes frequency changes that are strongly related to
the distribution in latitude of the magnetic field
\citep[e.g.,][]{Howe2002}.
As noted by \citet{Dziemb1997} the concentration
of magnetic activity towards the equator causes a frequency shift for
low-degree modes, observed with limited frequency resolution,
that depends on degree and hence might corrupt the study of the
solar core based, e.g., on the small frequency separations.
Such degree-dependent frequency shifts were in fact observed
by \citet{Chapli2004} and \citet{Toutai2005}.
It was argued by \citet{Dziemb1997} that the effects could be eliminated
in the solar case from observations of higher-degree modes;
however, it is obvious that they are a significant concern in
observations of distant stars where only low-degree data are available.

%\note [Hence discuss effect on frequencies, diagnostic power of 
%asteroseismology, with reference to the much more detailed presentation
%by Lebreton. (And with reference to the etymology of asteroseismology.)]
%
%\note [Fit small separations as Bonanno et al. (also check Dziembowski et al.)
%with new models, old and new composition.
%Note that \citet{Dziemb1999} checked the effect of changing $Z/X$.
%Use \citet{Iglesi1996} opacities and \citet{Rogers2002} EOS; otherwise, as
%Model S.]

The independent radioactive age determination of the solar system
provides an excellent test of the use of the oscillation frequencies
of low-degree acoustic modes to determine stellar ages.
\citet{Gough1990} made a careful analysis of the 
sensitivity of seismic age determinations to other aspects of the
solar models.
This was extended by \citet{Gough2001}, including also a determination
of $Z_{\rm s}/X_{\rm s}$, from the analysis of $\delta \nu_{02}$ and
$\delta \nu_{03}$;
the results were consistent with the meteoritic ages although with
a substantial uncertainty, owing to the strong sensitivity
of the small frequency spacings to $Z_{\rm s}/X_{\rm s}$.
\citet{Guenth1997} also estimated the solar age based on the
small frequency separations.  More systematic analyses, using a
$\chi^2$ fit to the observed values, fixing the value of $Z_{\rm
s}/X_{\rm s}$, were carried out by \citet{Dziemb1999} and
\citet{Bonann2002}.  These analyses showed that the seismically
inferred age was in good agreement with the meteoritic age.
Dziembowski et al.\ also found that the inferred age decreased with an
increased $Z_{\rm s}/X_{\rm s}$, in accordance with the results of
\citet{Gough2001}.

I have repeated this type of analysis, using the observed frequencies of
\citet{Chapli2007}, based on 4572 days of observation with the BiSON
network and corrected for frequency shifts caused by the solar
magnetic activity.  The models essentially corresponded to Model S of
\citet{Christ1996}, except that more recent OPAL opacities
\citep{Iglesi1996} and OPAL equation of state \citep{Rogers2002}%
\footnote{In particular, the equation of state tables included relativistic
effects for the electrons; \citet{Bonann2002} found that these had a noticeable
effect on the age fit.}
were used.
The goodness of fit was determined by, for example,
\begin{equation}
\chi^2(\delta \nu_{02}) = {1 \over N - 1} 
\sum_n {[\delta \nu_{02}(n)^{\rm (obs)} - \delta \nu_{02}(n)^{\rm (mod)}]^2
\over \sigma[\delta \nu_{02}(n)]^2} \; ,
\label{eq:chisq}
\end{equation}
where $N$ is the number of modes included, $\delta \nu_{02}(n)^{\rm
(obs)}$ and $\delta \nu_{02}(n)^{\rm (mod)}$ are the observed and
model values of the small separation, and $\sigma[\delta
\nu_{02}(n)]^2$ is the variance of the observed small separation.

\begin{figure}[b]
% \vspace*{-2.0 cm}
\begin{center}
 \includegraphics[width=11cm]{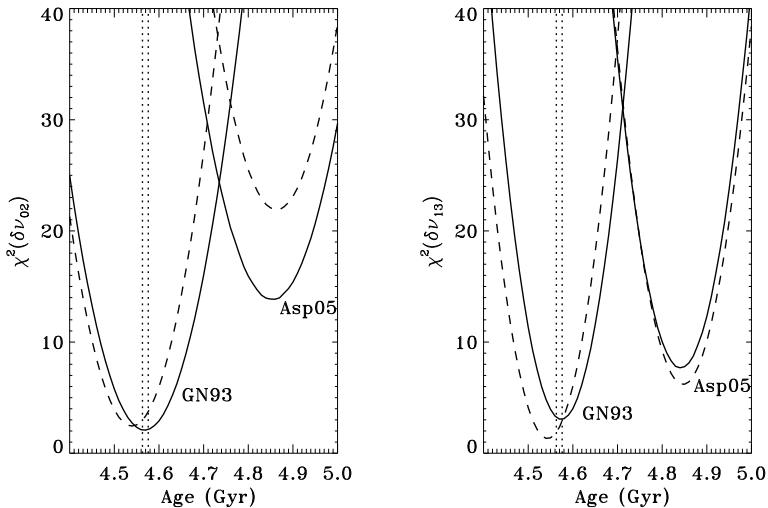} 
% \vspace*{-1.0 cm}
 \caption{Fits of solar models, as a function of age,
to the frequency separations obtained from BiSON observations
\citep{Chapli2007};
the left-hand panel shows results for the separations between $l = 0$
and 2, and the right-hand panel for separations between $l = 1$ and 3.
The solid curves are for $\delta \nu_{l\,l+2}$ and the dashed curves
for $r_{l\,l+2}$.  Results are shown both for models computed with the
\citet{Greves1993} composition (GN93) and the \citet{Asplun2005}
composition (Asp05).  The vertical dotted lines indicate the interval
of solar age obtained by Wasserburg, in \citet{Bahcal1995}.}
   \label{fig:chisq}
\end{center}
\end{figure}

Preliminary results of this analysis are shown in
Fig.~\ref{fig:chisq}.  Fits have been made to both the unscaled small
separations $\delta \nu_{02}$ and $\delta \nu_{13}$ and the separation
ratios $r_{02}$ and $r_{13}$.  I have computed results as functions of
age, in all cases calibrating the models to the solar radius and
luminosity and a fixed value of $Z_{\rm s}/X_{\rm s}$.  The left-hand
curves assumed $Z_{\rm s}/X_{\rm s} = 0.0245$
\citep{Greves1993};
here the best fits are evidently obtained close to the age interval
obtained from the meteoritic analysis, indicated by the vertical
lines.  Interestingly, the fit for the separation ratios indicate a
slightly lower age and, in the case of $r_{13}$, a substantially lower
$\chi^2$; in general, the values of $\chi^2$ show that the models are
not entirely consistent with the observations.  On the other hand, the
results in the right-hand curves for the \citet{Asplun2005}
composition, with $Z_{\rm s}/X_{\rm s} = 0.0165$, are clearly entirely
inconsistent with the meteoritic age.  This is in agreement with the
analysis by \citet{Chapli2007} of the small separations, similarly
showing that they are incompatible with the revised composition.
Also, the values of the minimal $\chi^2$ are much larger than for the
GN93 composition.  The very large minimum $\chi^2$ for $r_{02}$ in
this case clearly requires further investigation.  Note that the shift
in the inferred age with $Z_{\rm s}/X_{\rm s}$ is in accordance with
the results obtained by \citet{Dziemb1999} and \citet{Gough2001}.

As noted by \citet{Gough2002} the systematic errors, arising from the
other unknown parameters of the stars, far exceed the effects of the
statistical errors in the oscillation frequencies.  To constrain these
parameters on the basis of asteroseismic data he proposed the analysis
of other aspects of the frequencies, particularly the effects sharp
features in the sound speed (see also Lebreton \& Montalb\'an, this
volume). \citet{Houdek2007a} made a detailed analysis of the effects
of such features, which they called `acoustical glitches', in order to
derive reliable diagnostics for the envelope helium abundance and the
depth of the convection zone.  This was applied by \citet{Houdek2007b}
and \citet{Houdek2008} to the determination of the solar age, based
solely on low-degree frequencies such as might be observed in other
stars; they noted that removing the effects of the glitches from the
frequencies resulted in a more robust calibration for the age.  The
resulting age, calibrating the radius and luminosity to solar values
but determining the heavy-element abundance from the fit, is close to,
but not entirely consistent with, the meteoritic age; preliminary
results are an age $t_\odot = (4.68 \pm 0.02) \Gyr$ and an initial
heavy-element abundance $Z_0 = 0.0169 \pm 0.0005$.

%\note [Comment on effects of other uncertainties, still to be quantified.
%(Does RoxVor scaling properly correct for radius error? Deserves a check!)]
%
%\note [Mention briefly possibility of inversion in core, perhaps.]
%
%\note [Analysis of acoustic glitches, to correct for other uncertainties,
%as Houdek \& Gough.]
%
%\note [Summary of H\&G results.]

It is evident that processes modifying the core composition have a
potentially serious effect on the age determination.  This is
particularly true for stars with convective cores where the uncertain
extent of convective overshoot has a large effect on the relation
between age and stellar structure, as illustrated by Lebreton \&
Montalb\'an (this volume) in a $(\langle \Delta \nu \rangle, \langle
r_{02} \rangle)$ diagram.  To detect such effects and correct the age
determinations for them require asteroseismic analysis beyond the
simple fits to the frequency separations.  It is encouraging that it
appears to be possible, with sufficiently good data, to resolve the
structure of stellar cores in inverse analyses using just low-degree
modes \citep[e.g.,][]{Gough1993, Basu2002, Roxbur2002}.  This may
allow determination of the extent of convective overshoot and other
mixing processes that could affect the age determination.  Also, based
on asymptotic analysis \citet{Cunha2007} developed a diagnostic tool
which may be used to characterize small convective cores and hence
potentially eliminate the effects of additional mixing.  A similar
diagnostic was found by \citet{Mazumd2006} on the basis of extensive
model calculations.

\section{Next steps}

%\note [Prepare for coming observations. Quantify effects of other 
%uncertainties.]
%
We have yet to see the full realization of the potential for age
determination based on asteroseismology, but the observational
prospects are excellent.  The CoRoT mission has yielded the first
results on solar-like stars
\citep[e.g.,][]{Appour2008} and much more is expected in the
next few years.  The NASA Kepler mission, with planned launch in March
2009, will yield excellent asteroseismic data for a very large number
of stars \citep[e.g.,][]{Christ2007, Kjelds2008b}; an important aspect
of the asteroseismic investigation based on Kepler data, given the
main goal of the mission of characterizing extra-solar planetary
systems, is to determine properties of the central stars in such
systems, in particular their radius and age.  In the longer term,
ground-based projects for Doppler-velocity observations of stellar
oscillations, such as the SONG \citep{Grunda2008} and SIAMOIS
\citep{Mosser2008} projects, are expected to yield exquisite data
although for a smaller number of stars.

%\note [Goal: proper calibration of other age diagnostics.]
%
%\note [Coming observational projects, very briefly, with references.]

To utilize fully the data from these projects we need further
development and tests of the asteroseismic diagnostic tools, taking
into account also the additional unknown properties of the stars on
the one hand, and other observed properties of the stars on the other.
There is no doubt that, as in the past, asymptotic analyses will be
extremely important guides in determining the optimal combinations of
frequencies; however, extensive model calculations and analysis of
artificial data, under the various relevant assumptions, will also be
crucial.  This clearly needs to take into account also the detailed
properties of the oscillations and their effect on the inferred
oscillation parameters \citep[e.g.,][]{Chapli2008a}.  A central effort
in this regard is the asteroFLAG project
\citep{Chapli2008b, Chapli2008c} to carry out blind tests on
the analysis of artificial data, involving a substantial number of
different techniques and groups.

As a result of these efforts, both observational and theoretical, we
may hope to obtain reliable and precise age determinations for a
number of stars of varying properties.  These can then be used as
calibrators for other, less direct, age diagnostics and thus extend
the base for the general determination of stellar ages.

\begin{discussion}

\discuss{G. Meynet}{Can you say a few words about the behavior of the
angular velocity near the center of the Sun? Is it increasing or
decreasing towards the center? What are the most recent results?}

\discuss{J. Christensen-Dalsgaard}{The data are consistent with
constant rotation in the core. Unfortunately, improving the error bars
with p-mode observations will require very extended observations and
the g-mode claims, although very interesting, are so far tentative.}

\discuss{D. Soderblom}{You mentioned that Kepler may be able to detect
planets from phase shifts of oscillation frequencies. Has that effect
been seen on the Sun?}

\discuss{J. Christensen-Dalsgaard}{I suppose that the effect of
Jupiter might be visible, but it has not been seen, or looked for, as
far as I know.}

\discuss{S. Leggett}{Can you comment further on the Asplund
abundances? Have they been revised upward?}

\discuss{J. Christensen-Dalsgaard:}{There has been an independent
analysis of a similar nature by Caffau et al. (2008, A \&A, 488,
1031); preliminary results show an oxygen abundance halfway between
the old and the Asplund values.}

\discuss{P. Demarque}{I draw your attention to a recent detailed
review of the solar abundance problem by Basu \& Antia (2008,
Phys. Rep., 457, 217). There is also a poster downstairs in which my
collaborators and I point out some internal inconsistencies in the
Asplund et al. analysis. Having worked on both helioseismology and 3-D
simulations, I must say that I consider the seismic results to be more
trustworthy.}

\end{discussion}

%\begin{figure}[htbp]
%\begin{center}
%\includegraphics[angle=0,width=4in]{IAU258_hillenbrand.eps}\\
%Lynne Hillenbrand
%\end{center}
%\end{figure}
%
%\begin{figure}[htbp]
%\begin{center}
%\includegraphics[angle=0,width=4in]{IAU258_cruz_cody_frebel.eps}\\
%Ann Marie Cody, Kelle Cruz, and Anna Frebel
%\end{center}
%\end{figure}


\begin{thebibliography}{}

%\bibitem[Amari \etal\ (1995)]{Amari_etal95}
%{Amari, S., Hoppe, P., Zinner, E., \& Lewis R.S.} 1995,
%\textit{Meteoritics}, 30, 490 
%
%\bibitem[Anders \& Zinner (1993)]{AndersZinner93}
%{Anders, E., \& Zinner, E.} 1993, 
%\textit{Meteoritics}, 28, 490
%

\bibitem[Ahmad {\etal}(2002)]{Ahmad2002}
Ahmad, Q. R., Allen, R. C., Andersen, T. C., \etal\ 2002,
%Anglin, J. D., Barton, J. C.,
%Beier, E. W., Bercovitch, M., Bigu, J., Biller, S. D., Black, R. A.,
%Blevis, I., Boardman, R. J., Boger, J., Bonvin, E., Boulay, M. G.,
%Bowler, M. G., Bowles, T. J., Brice, S. J., Browne, M. C., Bullard, T. V.,
%B\"uhler, G., Cameron, J., Chan, Y. D., Chen, H. H., Chen, M., Chen, X.,
%Cleveland, B. T., Clifford, E. T. H., Cowan, J. H. M., Cowen, D. F.,
%Cox, G. A., Dai, X., Dalnoki-Veress, F., Davidson, W. F., Doe, P. J.,
%Doucas, G., Dragowsky, M. R., Duba, C. A., Duncan, F. A., Dunford, M.,
%Dunmore, J. A., Earle, E. D., Elliott, S. R., Evans, H. C., Ewan, G. T.,
%Farine, J., Fergani, H., Ferraris, A. P., Ford, R. J., Formaggio, J. A.,
%Fowler, M. M., Frame, K., Frank, E. D., Frati, W., Gagnon, N., Germani, J. V.,
%Gil, S., Graham, K., Grant, D. R., Hahn, R. L., Hallin, A. L., Hallman, E. D.,
%Hamer, A. S., Hamian, A. A., Handler, W. B., Haq, R. U., Hargrove, C. K.,
%Harvey, P. J., Hazama, R., Heeger, K. M., Heintzelman, W. J., Heise, J.,
%Helmer, R. L., Hepburn, J. D., Heron, H., Hewett, J., Hime, A., Howe, M.,
%Hykawy, J. G., Isaac, M. C. P., Jagam, P., Jelley, N. A., Jillings, C.,
%Jonkmans, G., Kazkaz, K., Keener, P. T., Klein, J. R., Knox, A. B.,
%Komar, R. J., Kouzes, R., Kutter, T., Kyba, C. C. M., Law, J., Lawson, I. T.,
%Lay, M., Lee, H. W., Lesko, K. T., Leslie, J. R., Levine, I., Locke, W.,
%Luoma, S., Lyon, J., Majerus, S., Mak, H. B., Maneira, J., Manor, J.,
%Marino, A. D., McCauley, N., McDonald, A. B., McDonald, D. S., McFarlane, K.,
%McGregor, G., Meijer Drees, R., Mifflin, C., Miller, G. G., Milton, G.,
%Moffat, B. A., Moorhead, M., Nally, C. W., Neubauer, M. S., Newcomer, F. M.,
%Ng, H. S., Noble, A. J., Norman, E. B., Novikov, V. M., O'Neill, M.,
%Okada, C. E., Ollerhead, R. W., Omori, M., Orrell, J. L., Oser, S. M.,
%Poon, A. W. P., Radcliffe, T. J., Roberge, A., Robertson, B. C.,
%Robertson, R. G. H., Rosendahl, S. S. E., Rowley, J. K., Rusu, V. L.,
%Saettler, E., Schaffer, K. K., Schwendener, M. H., Sch\"ulke, A., Seifert, H.,
%Shatkay, M., Simpson, J. J., Sims, C. J., Sinclair, D., Skensved, P.,
%Smith, A. R., Smith, M. W. E., Spreitzer, T., Starinsky, N., Steiger, T. D.,
%Stokstad, R. G., Stonehill, L. C., Storey, R. S., Sur, B., Tafirout, R.,
%Tagg, N., Tanner, N. W., Taplin, R. K., Thorman, M., Thornewell, P. M.,
%Trent, P. T., Tserkovnyak, Y. I., Van Berg, R., Van de Water, R. G.,
%Virtue, C. J., Waltham, C. E., Wang, J.-X., Wark, D. L.,
%West, N., Wilhelmy, J. B., Wilkerson, J. F., Wilson, J. R., Wittich, P.,
%Wouters, J. M. \& Yeh, M. 2002,
%[Direct evidence for neutrino flavor transformation from neutral-current
%interactions in the Sudbury Neutrino Observatory].
{\textit Phys. Rev. Lett.} {\rm 89}, 011301 %-(1 -- 6).

\bibitem[Appourchaux {\etal}(2008)]{Appour2008}
Appourchaux, T., Michel, E., Auvergne, M., \etal\ 2008,
%Baglin, A., Toutain, T., Baudin, F.,
%Benomar, O., Chaplin, W. J., Deheuvels, S., Samadi, R., Verner, G. A.,
%Boumier, P., Garc\'{\i}a, R. A., Mosser, B., Hulot, J.-C., Ballot, J.,
%Barban, C., Elsworth, Y., Jim\'enez-Reyes, S. J., Kjeldsen, H., 
%R\'egulo, C., Roxburgh, I. W. 2008,
%[CoRoT sounds the stars: p-mode parameters of Sun-like oscillations
%on HD 49933].
{\textit A\&A} {\rm 488}, 705 % -- 714.

\bibitem[Asplund(2005)]{Asplun2005}
Asplund, M. 2005,
%[New light on stellar abundance analysis: departures from LTE and 
%homogeneity].
{\textit ARAA} {\rm 43}, 481 % -- 540.

\bibitem[Asplund {\etal}(2004)]{Asplun2004}
Asplund, M., Grevesse, N., Sauval, A. J., Allende Prieto, C. \&
Kiselman, D. 2004,
%[Line formation in solar granulation. IV.
%[O I], O I and OH lines and the photospheric O abundance].
{\textit A\&A} {\rm 417}, 751 % -- 768
(Erratum: {\textit A\&A} {\rm 435}, 339) % -- 340).

\bibitem[Auwers(1891)]{Auwers1891}
Auwers, A. 1891,
%[Der Sonnendurchmesser und der Venusdurchmesser nach den Beobachtungen
%an den Heliometern der deutschen Venus-Expedition].
{\textit Astron. Nachr.} {\rm 128}, 361 %-- 375.

\bibitem[B\"ohm-Vitense(1958)]{Bohm1958}
B\"ohm-Vitense, E. 1958,
%[\"Uber die Wasserstoffkonvektionszone in Sternen
%verschiedener Effektivtemperaturen und Leuchtkr\"afte].
{\textit ZfA} {\rm 46}, 108 %-- 143.

\bibitem[Bahcall \& Pinsonneault(1995)]{Bahcal1995}
Bahcall, J. N. \& Pinsonneault, M. H. 1995,
(with an appendix by G. J. Wasserburg),
%[Solar models with helium and heavy-element diffusion].
{\textit Rev. Mod. Phys.} {\rm 67}, 781 %-- 808.

\bibitem[Bahcall \& Serenelli(2005)]{Bahcal2005a}
Bahcall, J. N. \& Serenelli, A. M. 2005,
%[How do uncertainties in the surface chemical composition of the Sun affect
%the predicted solar neutrino fluxes?].
{\textit ApJ} {\rm 626}, 530 %-- 542.

\bibitem[Bahcall {\etal}(2004)]{Bahcal2004}
Bahcall, J. N., Gonzalez-Garcia, M. C. \& Pe\~na-Garay, C. 2004,
%[Solar neutrinos before and after Neutrino 2004].
{\textit J. High Energy Phys.} {\rm 08}, 016 %-(1--26).

\bibitem[Bahcall {\etal}(2005)]{Bahcal2005b}
Bahcall, J. N., Basu, S., Pinsonneault, M. \& Serenelli, A. M. 2005,
%[Helioseismological implications of recent solar abundance determinations].
{\textit ApJ} {\rm 618}, 1049 %-- 1056. % {\tt [astro-ph/0407060v1]}.

\bibitem[Bahcall {\etal}(2006)]{Bahcal2006}
Bahcall, J. N., Serenelli, A. M. \& Basu, S. 2006,
%[10,000 standard solar models: a Monte Carlo simulation].
{\textit ApJS} {\rm 165}, 400 %-- 431.

\bibitem[Barnes(2007)]{Barnes2007}
Barnes, S. A. 2007,
%[Ages for illustrative field stars using gyrochronology: validity,
%limitations, and errors].
{\textit ApJ} {\rm 669}, 1167 %-- 1189.

\bibitem[Basu \& Antia(2008)]{Basu2008}
Basu, S. \& Antia, H. M. 2008,
%[Helioseismology and solar abundances].
{\textit Phys. Rep.} {\rm 457}, 217 %-- 283.

\bibitem[Basu {\etal}(2002)]{Basu2002}
Basu, S., Christensen-Dalsgaard, J. \& Thompson, M. J. 2002,
%[SOLA inversions for the core structure of solar-type stars].
in: F. Favata, I. W. Roxburgh \& D. Galad\'{\i}-Enr\'{\i}quezi (eds),
{\textit Proc.\ 1st Eddington Workshop, `Stellar Structure
and Habitable Planet Finding'},
ESA SP-485 (Noordwijk, The Netherlands: ESA),
p.\ 249 %-- 252.

\bibitem[Bonanno {\etal}(2002)]{Bonann2002}
Bonanno, A., Schlattl, H. \& Patern\`o, L. 2002,
%[The age of the Sun and the relativistic corrections in the EOS].
{\textit A\&A} {\rm 390}, 1115 %-- 1118.

\bibitem[Brown \& Christensen-Dalsgaard(1998)]{Brown1998}
Brown, T. M. \& Christensen-Dalsgaard, J. 1998,
%[Accurate determination of the solar photospheric radius].
{\textit ApJ} {\rm 500}, L195 %-- L198.

\bibitem[Chaplin {\etal}(2004)]{Chapli2004}
Chaplin, W. J., Elsworth, Y., Isaak, G. R., Miller, B. A. \& New, R. 2004,
%[The solar cycle as seen by low-$\ell$ p-mode frequencies: comparison with
%global and decomposed activity proxies].
{\textit MNRAS} {\rm 352}, 1102 %-- 1108.

\bibitem[Chaplin {\etal}(2007)]{Chapli2007}
Chaplin, W. J., Serenelli, A. M., Basu, S., Elsworth, Y., New, R. \&
Verner, G. A. 2007,
%[Solar heavy-element abundance: constraints from frequency separation ratios
%of low-degree $p$-modes].
{\textit ApJ} {\rm 670}, 872 %-- 884.

\bibitem[Chaplin {\etal}(2008a)]{Chapli2008a}
Chaplin, W. J., Houdek, G., Appourchaux, T., Elsworth, Y., New, R. \&
Toutain, T. 2008a,
%[Challenges for asteroseismic analysis of Sun-like stars].
{\textit A\&A} {\rm 485}, 813 %-- 822.

\bibitem[Chaplin {\etal}(2008b)]{Chapli2008b}
Chaplin, W. J., Appourchaux, T., Arentoft, T., \etal\ 2008b,
%Ballot, J.,
%Christensen-Dalsgaard, J., Creevey, O. L., Elsworth, Y., Fletcher, S. T.,
%Garc\'{\i}a, R. A., Houdek, G., Jim\'enez-Reyes, S. J., Kjeldsen, H.,
%New, R., R\'egulo, C., Salabert, D., Sekii, T., Sousa, S. G., Toutain, T.,
%and the rest of the asteroFLAG group, 2008b.
%[AsteroFLAG: First results from hare-and-hound Exercise \#1].
{\textit Astron. Nach.} {\rm 329}, 549 %-- 557.

\bibitem[Chaplin {\etal}(2008c)]{Chapli2008c}
Chaplin, W. J., Appourchaux, T., Arentoft, T., \etal\ 2008c,
%Ballot, J., Baudin, F.,
%Bazot, M., Bedding, T. R., Christensen-Dalsgaard, J., Creevey, O. L., Duez, V.,
%Elsworth, Y., Fletcher, S. T., Garc\'{\i}a, R. A., Gough, D. O.,
%Jim\'enez, A., Jim\'enez-Reyes, S. J., Houdek, G., Kjeldsen, H.,
%Lazrek, M., Leibacher, J. W., Monteiro, M. J. P. F. G., Neiner, C.,
%New, R., R\'egulo, C., Salabert, D., Samadi, R., Sekii, T., Sousa, S. G.,
%Toutain, T. \& Turck-Chi\`eze, S., 2008c.
%[AsteroFLAG -- from the Sun to the stars].
in: L. Gizon \& M. Roth (eds),
{\textit Proc.\ HELAS II International Conference: Helioseismology,
Asteroseismology and the MHD Connections},
{\textit J. Phys.: Conf. Ser.} {\rm 118}, 012048 %(1 -- 5).

\bibitem[Christensen-Dalsgaard(1984)]{Christ1984}
Christensen-Dalsgaard, J. 1984,
%[What will asteroseismology teach us?].
in: A. Mangeney \& F. Praderie (eds),
{\textit Space Research Prospects in Stellar Activity and Variability}
(Paris: Paris Observatory Press),
p.\ 11 %-- 45.

\bibitem[Christensen-Dalsgaard(1988)]{Christ1988b}
Christensen-Dalsgaard, J. 1988,
%[A Hertzsprung-Russell diagram for stellar oscillations].
in: J. Christensen-Dalsgaard \& S. Frandsen (eds), 
{\textit Advances in helio- and asteroseismology},
Proc.\ IAU Symposium No 123
(Dordrecht: Reidel), p.\ 295 %-- 298.

\bibitem[Christensen-Dalsgaard(2002)]{Christ2002}
Christensen-Dalsgaard, J. 2002,
%[Helioseismology].
{\textit Rev. Mod. Phys.} {\rm 74}, 1073 %-- 1129. % Oct. 2002, [astro-ph/0207403].

\bibitem[Christensen-Dalsgaard \& Thompson(1997)]{Christ1997}
Christensen-Dalsgaard, J. \& Thompson, M. J. 1997,
%[On solar p-mode frequency shifts caused by near-surface model changes].
{\textit MNRAS} {\rm 284}, 527 %-- 540.

\bibitem[Christensen-Dalsgaard {\etal}(1988)]{Christ1988a}
Christensen-Dalsgaard, J., D\"appen, W. \& Lebreton, Y. 1988,
%[Solar oscillation frequencies and the equation of state].
{\textit Nature} {\rm 336}, 634 %-- 638.

\bibitem[Christensen-Dalsgaard {\etal}(1996)]{Christ1996}
Christensen-Dalsgaard, J., D\"appen, W., Ajukov, S. V., \etal\ 1996,
%Anderson, E. R.,
%Antia, H. M., Basu, S., Baturin, V. A., Berthomieu, G., Chaboyer, B.,
%Chitre, S. M., Cox, A. N., Demarque, P., Donatowicz, J., Dziembowski, W. A.,
%Gabriel, M., Gough, D. O., Guenther, D. B., Guzik, J. A., Harvey, J. W.,
%Hill, F., Houdek, G., Iglesias, C. A., Kosovichev, A. G., Leibacher, J. W.,
%Morel, P., Proffitt, C. R., Provost, J., Reiter, J., Rhodes Jr., E. J.,
%Rogers, F. J., Roxburgh, I. W., Thompson, M. J. \& Ulrich, R. K. 1996,
%[The current state of solar modeling].
{\textit Science} {\rm 272}, 1286 %-- 1292.

\bibitem[Christensen-Dalsgaard {\etal}(2007)]{Christ2007}
Christensen-Dalsgaard, J., Arentoft, T., Brown, T. M., Gilliland, R. L.,
Kjeldsen, H., Borucki, W. J. \& Koch, D. 2007,
%[Asteroseismology with the {\em Kepler mission\/}].
in: G. Handler \& G. Houdek (eds),
{\textit Proc.\ Vienna Workshop on the Future of Asteroseismology},
{\textit Comm. in Asteroseismology} {\rm 150}, 350 %-- 356.

\bibitem[Christensen-Dalsgaard {\etal}(2009)]{Christ2009}
Christensen-Dalsgaard, J., Di Mauro, M. P., Houdek, G. \& Pijpers, F. 2009,
%[On the opacity change required to compensate for the revised solar
%composition].
{\textit A\&A}, in the press {\tt [arXiv:0811.1001 [astro-ph]]}.

\bibitem[Cunha \& Metcalfe(2007)]{Cunha2007}
Cunha, M. S. \& Metcalfe, T. S. 2007,
%[Asteroseismic signatures of small convective cores].
{\textit ApJ} {\rm 666}, 413 %-- 422.

\bibitem[Demarque {\etal}(1999)]{Demarq1999}
Demarque, P., Guenther, D. B. \& Kim, Y.-C. 1999,
%[The run of superadiabaticity in stellar convection zones. II.
%Effect of photospheric convection on solar $p$-mode frequencies].
{\textit ApJ} {\rm 517}, 510 %-- 515.

\bibitem[Dziembowski \& Goode(1997)]{Dziemb1997}
Dziembowski, W. A. \& Goode, P. R. 1997,
%[Seismic sounding of the solar core: purging the corruption 
%from the Sun's magnetic field].
{\textit A\&A} {\rm 317}, 919 %-- 924.

\bibitem[Dziembowski {\etal}(1999)]{Dziemb1999}
Dziembowski, W. A., Fiorentini, G., Ricci, B. \& Sienkiewicz, R. 1999,
%[Helioseismology and the solar age].
{\textit A\&A} {\rm 343}, 990 %-- 996.

\bibitem[Elsworth {\etal}(1990)]{Elswor1990}
Elsworth, Y., Howe, R., Isaak, G. R., McLeod, C. P. \& New, R. 1990,
%[Evidence from solar seismology against non-standard solar-core models].
{\textit Nature} {\rm 347}, 536 %-- 539.

\bibitem[Fr\"ohlich \& Lean(2004)]{Frohli2004}
Fr\"ohlich, C. \& Lean, J. 2004,
%[Solar radiative output and its variability: evidence and mechanisms].
{\textit A\&AR} {\rm 12}, 273 %-- 320.

\bibitem[Gough(1987)]{Gough1987}
Gough, D. O. 1987,
%[Seismological measurement of stellar ages].
{\textit Nature} {\rm 326}, 257 %-- 259.

\bibitem[Gough(1996)]{Gough1996}
Gough, D. O. 1996,
%[Astereoasteroseismology (correspondence to the Editors)].
{\textit Observatory} {\rm 116}, 313 %-- 315.

\bibitem[Gough(2001)]{Gough2001}
Gough, D. O. 2001,
%[Lessons learned from solar oscillations].
in: T. von Hippel, C. Simpson \& N. Manset (eds),
{\textit Astrophysical Ages and Time Scales},
ASP Conf. Ser. {\rm 245} 
(San Francisco: ASP), p.\ 31 %-- 43.

\bibitem[Gough(2002)]{Gough2002}
Gough, D. O. 2002,
%[Helioseismology: some current issues concerning model calibration].
in: F. Favata, I. W. Roxburgh \& D. Galad\'{\i}-Enr\'{\i}quezi (eds),
{\textit Proc.\ 1st Eddington Workshop: `Stellar structure and habitable
planet finding'}, 
ESA SP-485
(Noordwijk, The Netherlands: ESA),
p.\ 65 %-- 73.

\bibitem[Gough \& Kosovichev(1993)]{Gough1993}
Gough, D. O. \& Kosovichev, A. G. 1993,
%[Seismic analysis of stellar p-mode spectra].
in:
T. M. Brown (ed.),
{\textit Proc.\ GONG 1992: Seismic investigation of the Sun and stars},
ASP Conf. Ser. {\rm 42}, (San Francisco: ASP), p.\ 351 %-- 354.

\bibitem[Gough \& Novotny(1990)]{Gough1990}
Gough, D. O. \& Novotny, E. 1990,
%[Sensitivity of solar eigenfrequencies to the age of the Sun].
{\textit Solar Phys.} {\rm 128}, 143 %-- 160.

\bibitem[Gough \& Weiss(1976)]{Gough1976}
Gough, D. O. \& Weiss, N. O. 1976,
%[The calibration of stellar convection theories].
{\textit MNRAS} {\rm 176}, 589 %-- 607.

\bibitem[Grevesse \& Noels(1993)]{Greves1993}
Grevesse, N. \& Noels, A. 1993,
%[Cosmic abundances of the elements].
in: N. Prantzos, E. Vangioni-Flam \& M. Cass\'e (eds),
{\textit Origin and evolution of the Elements},
(Cambridge: Cambridge Univ. Press), p.\ 15 %-- 25.

\bibitem[Grundahl {\etal}(2008)]{Grunda2008}
Grundahl, F., Arentoft, T., Christensen-Dalsgaard, J., Frandsen, S.,
Kjeldsen, H. \& Rasmussen, P. K. 2008,
%[Stellar Oscillations Network Group -- SONG].
in: L. Gizon \& M. Roth (eds),
{\textit Proc.\ HELAS II International Conference: Helioseismology,
Asteroseismology and the MHD Connections},
{\textit J. Phys.: Conf. Ser.} {\rm 118}, 012041 %(1 -- 7).

\bibitem[Guenther \& Demarque(1997)]{Guenth1997}
Guenther, D. B. \& Demarque, P. 1997,
%[Seismic tests of the Sun's interior structure, composition, and age,
%and implications for solar neutrinos].
{\textit ApJ} {\rm 484}, 937 %-- 959.

%\bibitem[Guzik \& Cox(1991)]{Guzik1991}
%Guzik, J. A. \& Cox, A. N. 1991,
%%[Effects of opacity and equation of state on solar structure
%and oscillations].
%{\textitApJ} {\rm 381}, 333 %-- 340.

\bibitem[Guzik(2006)]{Guzik2006}
Guzik, J. A. 2006,
%[Reconciling the revised solar abundances with helioseismic constraints].
in: K. Fletcher (ed.),
{\textit Proc.\ SOHO 18 / GONG 2006 / HELAS I Conf.  Beyond the spherical Sun},
ESA SP-624, (Noordwijk, The Netherlands: ESA).

\bibitem[Guzik(2008)]{Guzik2008}
Guzik, J. A. 2008,
%[Problems for the standard solar model arising from the new solar mixture].
{\textit MemSAI} {\rm 79}, 481 %-- 489.

\bibitem[Haxton(2008)]{Haxton2008}
Haxton, W. C. 2008,
%[Solar neutrinos: models, observations and new opportunities].
{\textit PASA} {\rm 25}, 44 %-- 51.

\bibitem[Haxton {\etal}(2006)]{Haxton2006}
Haxton, W. C., Parker, P. D. \& Rolfs, C. E. 2006,
%[Solar hydrogen burning and neutrinos].
{\textit Nucl. Phys. A.} {\rm 777}, 226 %-- 253.

\bibitem[Houdek \& Gough(2007a)]{Houdek2007a}
Houdek, G. \& Gough, D. O. 2007a,
%[An asteroseismic signature of helium ionization].
{\textit MNRAS} {\rm 375}, 861 %-- 880.

\bibitem[Houdek \& Gough(2007b)]{Houdek2007b}
Houdek, G. \& Gough, D. O. 2007b,  
%[On the seismic age of the Sun].
in: R. J. Stancliffe, J. Dewi, G. Houdek, R. G. Martin \& C. A Tout (eds),
{\textit Unsolved Problems in Stellar Physics},
AIP Conf.\ Proc.\ 948 (Melville: AIP), p.\ 219 %-- 224.

\bibitem[Houdek \& Gough(2008)]{Houdek2008}
Houdek, G. \& Gough, D. O. 2008,  
%[Progress report on solar age calibration].
in: L. Deng \& K. L. Chan (eds),
{\textit The Art of Modelling Stars in the $21^{\rm st}$ Century},
Proc.\ IAU Symposium No 252 
(Cambridge: Cambridge University Press), p.\ 149 %-- 156.

\bibitem[Howe {\etal}(2002)]{Howe2002}
Howe, R., Komm, R. W. \& Hill, F. 2002,
%[Localizing the solar cycle frequency shifts in global $p$-modes].
{\textit ApJ} {\rm 580}, 1172 %-- 1187.

\bibitem[Iglesias \& Rogers(1996)]{Iglesi1996}
Iglesias, C. A. \& Rogers, F. J. 1996,
%[Updated OPAL opacities].
{\textit ApJ} {\rm 464}, 943 %-- 953.

\bibitem[Kjeldsen {\etal}(2008a)]{Kjelds2008a}
Kjeldsen, H., Bedding, T. R. \& Christensen-Dalsgaard, J. 2008a,
%[Correcting stellar oscillation frequencies for near-surface effects].
{\textit ApJ} {\rm 683}, L175 %-- L178. 

\bibitem[Kjeldsen {\etal}(2008b)]{Kjelds2008b}
Kjeldsen, H., Bedding, T. R. \& Christensen-Dalsgaard, J. 2008b,
%[Measurements of stellar properties through asteroseismology: a tool for
%planet transit studies].
in: F. Pont, D. Queloz \& D. D. Sasselov (eds),
{\textit Transiting Planets},
Proc.\ IAU Symposium No 253
(Cambridge: Cambridge University Press), in the press
{\tt [arXiv:0807.0508v1 [astro-ph]]}

\bibitem[Ludwig {\etal}(1997)]{Ludwig1997}
Ludwig, H.-G., Freytag, B. \& Steffen, M. 1997,
%[A calibration of mixing length theory based on RHD simulations
%of solar-type convection].
in: F. P. Pijpers, J. Christensen-Dalsgaard \& C. S. Rosenthal, C. S. (eds), 
{\textit SCORe'96: Solar Convection and Oscillations and their Relationship},
(Dordrecht: Kluwer), p.\ 59 %-- 64.

\bibitem[Ludwig {\etal}(1999)]{Ludwig1999}
Ludwig, H.-G., Freytag, B. \& Steffen, M. 1999,
%[A calibration of the mixing-length for solar-type stars
%based on hydrodynamical simulations. I. Methodological aspects and
%results for solar metallicity].
{\textit A\&A} {\rm 346}, 111 %-- 124.

\bibitem[Mazumdar {\etal}(2006)]{Mazumd2006}
Mazumdar, A., Basu, S., Collier, B. L. \& Demarque, P. 2006,
%[Asteroseismic diagnostics of stellar convective cores].
{\textit MNRAS} {\rm 372}, 949 %-- 958.

\bibitem[Monteiro {\etal}(2002)]{Montei2002}
Monteiro, M. J. P. F. G., Christensen-Dalsgaard, J. \&
Thompson, M. J. 2002,
%[Asteroseismic Inference for Solar-Type Stars].
in: F. Favata, I. W. Roxburgh \& D. Galad\'{\i}-Enr\'{\i}quezi (eds),
{\textit Proc.\ 1st Eddington Workshop: `Stellar structure and habitable
planet finding'}, 
ESA SP-485 (Noordwijk, The Netherlands: ESA),
p.\ 291 %-- 298.

\bibitem[Mosser {\etal}(2008)]{Mosser2008}
Mosser, B., Appourchaux, T., Catala, C., Buey, J.-T. and the 
SIAMOIS team 2008,
%[SIAMOIS: Seismic Interferometer to Measure Oscillations in the
%Interior of Stars].
in: L. Gizon \& M. Roth (eds),
{\textit Proc.\ HELAS II International Conference: Helioseismology,
Asteroseismology and the MHD Connections},
{\textit J. Phys.: Conf. Ser.} {\rm 118}, 012042 %(1 -- 6).

\bibitem[Ot\'{\i} Floranes {\etal}(2005)]{OtiFlo2005}
Ot\'{\i} Floranes, H., Christensen-Dalsgaard, J. \& Thompson, M. J. 2005,
%[The use of frequency-separation ratios for asteroseismology].
{\textit MNRAS} {\rm 356}, 671 %-- 679.

\bibitem[Robertson(2006)]{Robert2006}
Robertson, R.~G.~H. 2006,
%[Solar neutrinos].
{\textit Prog. Particle Nuclear Phys.} {\rm 57}, 90 %-- 105.

\bibitem[Rogers \& Nayfonov(2002)]{Rogers2002}
Rogers, F. J. \& Nayfonov, A. 2002,
%[Updated and expanded OPAL equation-of-state tables: implications 
%for helioseismology].
{\textit ApJ} {\rm 576}, 1064 %-- 1074.

\bibitem[Rosenthal {\etal}(1999)]{Rosent1999}
Rosenthal, C. S., Christensen-Dalsgaard, J., Nordlund, {\AA}.,
Stein, R. F. \& Trampedach, R. 1999,
%[Convective contributions to the frequencies of solar oscillations].
{\textit A\&A} {\rm 351}, 689 %-- 700.

\bibitem[Roxburgh \& Vorontsov(2003)]{Roxbur2003}
Roxburgh, I. W. \& Vorontsov, S. V. 2003,
%[The ratio of small to large separations of acoustic oscillations as 
%a diagnostic of the interior of solar-like stars].
{\textit A\&A} {\rm 411}, 215 %-- 220.

\bibitem[Roxburgh(2002)]{Roxbur2002}
Roxburgh, I. W. 2002,
%[The tools of asteroseismology].
in: F. Favata, I. W. Roxburgh \& D. Galad\'{\i}-Enr\'{\i}quezi (eds),
{\textit Proc.\ 1st Eddington Workshop: `Stellar structure and habitable
planet finding'}, 
ESA SP-485 (Noordwijk, The Netherlands: ESA),
p.\ 75 %-- 85.

\bibitem[Sackmann \& Boothroyd(2003)]{Sackma2003}
Sackmann, I.-Juliana \& Boothroyd, A. I. 2003,
%[Our Sun. V. A bright young Sun consistent with helioseismology and
%warm temperatures on ancient Earth and Mars].
{\textit ApJ} {\rm 583}, 1024 %-- 1039.

\bibitem[Thompson {\etal}(2003)]{Thomps2003}
Thompson, M. J., Christensen-Dalsgaard, J., Miesch, M. S. \& Toomre, J. 2003,
%[The internal rotation of the Sun].
{\textit ARAA} {\rm 41}, 599 %-- 643.

\bibitem[Toutain \& Kosovichev(2005)]{Toutai2005}
Toutain, T. \& Kosovichev, A. G. 2005,
%[Study of the solar cycle dependence of low-degree $p$-modes with
%Michelson Doppler Imager and VIRGO].
{\textit ApJ} {\rm 622}, 1314 %-- 1319.

\bibitem[Trampedach {\etal}(1999)]{Trampe1999}
Trampedach, R., Stein, R. F., Christensen-Dalsgaard, J. \&
Nordlund, {\textit A\&A}. 1999,
%[Stellar evolution with a variable mixing-length parameter].
in:
A. Gim\'enez, E.F. Guinan \& B. Montesinos (eds),
{\textit Theory and Tests of Convection in Stellar Structure},
ASP Conf. Ser. {\rm 173} (San Francisco: ASP), p.\ 233 %-- 236.

\bibitem[Ulrich(1986)]{Ulrich1986}
Ulrich, R. K. 1986,
%[Determination of stellar ages from asteroseismology].
{\textit ApJ} {\rm 306}, L37 %-- L40.

\end{thebibliography}
\end{document}